\documentclass{aa}
\usepackage{graphics}
\usepackage{graphicx}
\DeclareGraphicsExtensions{.pdf,.png,.jpg}
\usepackage{amssymb,amsmath,amsfonts}
\usepackage{natbib}
\usepackage{tikz}
\usepackage{tikz-3dplot}
\usepackage{paralist}
\usepackage[varg]{txfonts}
\usepackage{hyperref}
\usepackage{color}
\usepackage{mathrsfs}
\usepackage{txfonts}

\begin{document}

\newcommand{\myletter}{\mathscr{Q}}

\title{Into the third dimension: stochastic measurements of Stokes parameters within the Poincar\'e sphere}
\author{Jason L. Quinn}

\institute{2321 25th St. S. Unit 108, Arlington, VA 22206, USA, \email{jason.lee.quinn@gmail.com}
          }

\date{Received XXX XX, 20XX; accepted XXX XX, 20XX}

\abstract{Inspired by recent use of polarimetry to study the Cosmic Microwave Background
and extragalatic supernovae, a foray into the statistical properties of Stokes parameters
expressed in spherical coordinates is began, allowing circular polarization
and linear polarization to be treated in a unified manner. The use of
spherical coordinates is quite necessary as it permits a Stokes polarization
state to be expressed in terms of the customary polarization angles and degree of polarization
usually needed for human interpretation. As shall be demonstrated, circular and linear polarization
are not statistically independent quantities but intertwined in a way that is especially
important, for instance, at low signal-to-noise.
New distributions, classical estimators, and marginalizations are presented for this
``three-dimensional'' polarization problem including a generalization of the Rice distribution.
The paper concludes with discussion regarding the potential pitfalls of a lower
dimensional analysis.}


\maketitle

\section{Introduction}

Astronomical polarimetry and spectropolarimetry, which have historically
been under-utilized, are now routinely being used in
many experiments to search for and test new physics.
Besides the diverse application of polarimetric techniques in solar
and stellar astronomy, two areas of particular note are polarimetric observations
of the Cosmic Microwave Background (CMB) and spectropolarimetric
observations of supernovae (SNe).

Detection of inflation
signatures on the CMB using polarimetric imaging were
recently announced by the BICEP2 team \citep{2014PhRvL.112x1101A,2014ApJ...792...62A}.
This result would confirm
one of astrophysics most cherished but experimentally elusive
theories. After a review of the public BICEP2 data, some
have questioned the interpretation of the BICEP2 data and suggested that the foreground dust
contamination had been underestimated or that the signal
could have been produced by other mechanisms such as ``radio loops'' \citep{2014ApJ...789L..29L}. The BICEP2 team have stood by their results although they have moderated their language
in the peer-reviewed version of their paper \citep{2014PhRvL.112x1101A}. If valid, the BICEP2
discovery continues the trend set by all-sky mapping satellites
like COBE, WMAP, and Planck and various ground and balloon surveys
of the CMB being perhaps the single most fruitful source to test
cosmological theory and sometimes even fundamental physics.
Further results related to CMB polarization are expected in the
near future from BICEP3 \citep{2013IAUS..288...80K}, the Keck Array \citep{2012JLTP..167..827S}, POLARBEAR \citep{2014arXiv1403.2369T}, and the
Planck team \citep{2014arXiv1405.0871P}. It is clear that the CMB will continue to
be a vital source of cutting edge science; but, as the controversy
around the BICEP2 announcement indicates, polarization work is tricky and extreme
care must be made to handle properly statistical and
systematic error in the data.

Other research groups have been using spectropolarimetric
observations to investigate the systematic behavior and statistical
variation of thermonuclear (Type Ia) and core-collapse SNe (all other types)
arising from asymmetries. (The \citet{2008ARA&A..46..433W} Annual Review
is a good introduction.) Spectropolarimetry
is the only known way to probe asymmetries of unresolved sources
such as extra-galactic supernovae so it will continue to be an effective research tool.
The detection of asymmetries has helped place much needed constraints on
SNe explosion physics/progenitor scenarios (recent examples include \citet{2010ApJ...722.1162M,2012ApJ...754...63T,2012A&A...545A...7P,2013AJ....145...27Z,2013MNRAS.433L..20M})
and environments \citep{2014MNRAS.442.1166M}. Asymmetries of Type Ia SNe are of particular interest as high-redshift
events were famously used to discover cosmic acceleration \citep{1998AJ....116.1009R,1999ApJ...517..565P}, a result for which leads of two large teams shared the
2011 Nobel Prize. This acceleration implies the existence
of so-called ``dark energy'', whose physical nature is
completely puzzling and is perhaps the biggest
unsolved problem in astrophysics. As such,
any possible correlations between metrics of asymmetry in Type Ia SNe
and other observables \citep{2005ApJ...632..450L,2007Sci...315..212W,2010ApJ...725L.167M}
are of high importance because correlations may be used to identify contaminating Type Ia sub-classes \citep{2013Sci...340..170W}
or apply statistical corrections to the maximum brightness,
both of which could help refine the measurement
of the cosmological constant \citep{2011MNRAS.413.3075M}. Spectropolarimetry
of SNe is a blooming field but, just as with the
CMB, handling the statistical and systematic error of polarimetric
data for SNe is challenging.

As both polarimetric imaging and spectropolarimetry require dividing incoming light into bins, researchers in these fields often find themselves in photon-starved
situations. Indeed, in both subjects discussed above, photon-limited data at
low signal-to-noise must be confronted.
(Low signal-to-noise is often the hallmark of forefront science by its very nature.) This
problem is compounded in polarimetric imaging when a narrow-band
filter is needed. Working at low signal-to-noise will continue
to be an issue for polarimetrists.

Polarization data is often measured in Stokes parameters but
expressed in human-digestible quantities like degrees of polarization
and angles of polarization. Despite this, there has been relatively little attention given
to the surprisingly complicated statistics involved 
when making transformations to these quantities at low signal-to-noise.
Most of the published literature of this fundamental,
practical topic has focused on the analysis of linear polarization
alone that involves marginalizations over a two-dimensional
gaussian distribution over the $Q$-$U$ plane cross section of the
Poincar\'e sphere (the ``Poincar\'e disk'' is a useful expression although
it is already in use in other mathematical contexts). Marginalization over linear degree of polarization
results in the Rice distribution \citep{1945BSTJ...24...46R} while marginalization over the linear
polarization angle results in another distribution given by \citet{1965AnAp...28..412V}. The authors refer
to this marginalized approach as the ``one-dimensional problem''. A small number of papers
such as \citet{1997ApJ...476L..27W}, \citet{2006PASP..118.1340V}, and
\citet{2012A&A...538A..65Q} have used a non-marginalized distribution over
the Poincar\'e disk (the ``two-dimensional problem''), with the latter
being the first to put the subject on a rigorous
mathematical footing. To date, there has been 
little to no attention paid to the three-dimensional problem defined
by a three-dimensional distribution over the Poincar\'e sphere.
Progress on higher dimensional generalizations of the two-dimensional
problem have been made, for instance, by members of
the Planck Collaboration (\cite{2014arXiv1405.0871P,2014arXiv1406.6536M,2014arXiv1407.0178M})
by including intensity in the analysis but researchers have thus far
generally ignored circular polarization
when measuring linear polarization in astronomical sources (and vice versa);
however, the measurement of one is not actually independent of the other. This 
effect is generally negligible except when
one is working at very small signal-to-noise ratios
or with objects with polarization near 100\% (see \citet{2012A&A...538A..65Q}
for details on how large polarization introduces complication). The
aim of this paper is to elucidate some of this interdependence and to
see how it relates to the more common ways of treating linear
polarization measurements at low signal-to-noise via the
Rice distribution or through Bayesian techniques as outlined
in \citet{2006PASP..118.1340V} and \citet{2012A&A...538A..65Q}. The
results should be of interest to all polarimetrists and those
those studying the CMB or SNe in particular.

\section{The sampling distribution}

A very practical way of encoding the polarization state of
electromagnetic radiation is the Stokes parameter formalism
invented by George Stokes in \citeyear{stokes1852}. Stokes parameters consist of
four quantities: $I$ (intensity), $Q$ and $U$ (related to linear polarization),
and $V$ (related to circular polarization).\footnote{This notation is fairly common in
optical astronomy but other notations are also
in use in different disciplines like solar astronomy or optical physics.} They
are convenient quantities to measure but it is
easier for people to think about polarization in terms of percentages
and angles. This may be done by introducing a spherical
coordinate system $(p,\theta,\phi)$ on points of the Poincar\'e sphere (a
unit ball in Cartesian coordinates centered on the origin of $\frac{Q}{I_0}$, $\frac{U}{I_0}$, and $\frac{V}{I_0}$ axes),
which represents all possible polarization states.
Accessible modern introductions
to Stokes parameters and their theory are given in
\citet{2003isp..book.....D} and \citet{2002apsp.conf....1L}.

If an astronomical source's polarization state is measured with
gaussian error $(\Sigma_I,\Sigma_Q,\Sigma_U,\Sigma_V)$ on each Stokes parameter $(I,Q,U,V)$,
how do the corresponding spherical coordinates $(p,\theta,\phi)$
relate to the ``true'' polarization state $(p_0,\theta_0,\phi_0)$
determined uniquely from $(I_0,Q_0,U_0,V_0)$?
(Unsubscripted variables will usually refer to measured values
while ``0''-subscripted values will refer to ``true'' values.
Later, an overbar on some variables and distributions will indicate signal-to-noise
related quantities.)
As it turns out, a spherical coordinate system introduces non-trivial
biases into the measured quantities and the measurement itself
must be ``corrected'' to remove them. 
This paper extends \citet{2012A&A...538A..65Q}, which studied
these effects in polar coordinates before discussing
the Bayesian approach, to spherical coordinates. The reader is referred
to that paper for a more extensive introduction on Stokes
parameters.

Under the assumptions that the source intensity is large
compared to the measurement error, i.e., $I_0 \gg \Sigma_I$ which implies $I \approx I_0$ \citep{2012A&A...538A..65Q}\footnote{This somewhat technical assumption does not (especially at low signal-to-noise) imply $Q \approx Q_0$, $U \approx U_0$, or $V \approx V_0$. It is a weaker condition used implicitly (but without recognition) by many important past papers on polarization statistics even though it is crucial for defining the reduced Stokes parameters as, for instance, $q \equiv Q/I_0$ rather than the problematic $q \equiv Q/I$. Even more fundamentally, without the condition, studying a distribution over $I$, $Q$, $U$, and $V$ instead of just $Q$, $U$, and $V$ would be thrust upon us. That $I \approx I_0$ does not imply the other approximations is easy to understand intuitively with back-of-the-envelope calculations by comparing the relative (Poisson) error on the intensity to the relative expected (Poisson) error on the Stokes parameters for some low degree of polarization source and some assumed moderate number of total counts for the measured intensity (like, for instance, a 1\% source and $1000$ total counts). The cited paper details the proper role of the condition in the theory.}, and the error on a Stokes measurement is described by a three-dimensional
uncorrelated ellipsoidal Gaussian distribution (with semi-major axes $\Sigma_Q$, $\Sigma_U$, $\Sigma_V$), the normalized distribution, $F_C$, of the measured Stokes parameters ($Q$, $U$, $V$) given the ``true'' Stokes parameters ($Q_0$, $U_0$, $V_0$) is
\begin{equation}
\begin{split}
F_C(Q,U,V|Q_0,U_0,{}&V_0,\Sigma_Q,\Sigma_U,\Sigma_V) = \\
&\frac{1}{(2 \pi)^{3/2} \Sigma_Q \Sigma_U \Sigma_V} 
e^{-\left( \frac{(Q-Q_0)^2}{2 \Sigma_Q^2} + \frac{(U-U_0)^2}{2 \Sigma_U^2} + \frac{(V-V_0)^2}{2 \Sigma_V^2} \right)}.
\end{split}
\label{starteq}
\end{equation}
These parameters are not usually worked with directly. Instead ``normalized Stokes parameters''
are defined as $q \equiv Q/I_0$, $u \equiv U/I_0$, $v \equiv V/I_0$
(with also $q_0 \equiv Q_0/I_0$, $u_0 \equiv U_0/I_0$, and $v_0 \equiv V_0/I_0$),
and $\sigma_q \equiv \Sigma_Q/I_0$, $\sigma_u \equiv \Sigma_U/I_0$, and $\sigma_v \equiv \Sigma_V/I_0$, yielding a new normalized distribution
\begin{equation}
\begin{split}
f_C(q,u,v|q_0,u_0,v_0,\sigma_q,{}&\sigma_u,\sigma_v) = \\
&\frac{1}{(2 \pi)^{3/2} \sigma_q \sigma_u \sigma_v}
e^{-\left(\frac{(q-q_0)^2}{2 \sigma_q^2} + \frac{(u-u_0)^2}{2 \sigma_u^2} + \frac{(v-v_0)^2}{2 \sigma_v^2}\right)}.
\end{split}
\end{equation}
Lastly, one most often works with ``signal-to-noise'' versions of the Stokes parameters
defined as
$\overline{q} \equiv Q/\Sigma_Q =q/\sigma_q$,
$\overline{u} \equiv U/\Sigma_U =u/\sigma_u$, and
$\overline{v} \equiv V/\Sigma_V =v/\sigma_v$ which gives another distribution
\begin{equation}
\begin{split}
\overline{f}_C(\overline{q},\overline{u},\overline{v}|\overline{q}_0,\overline{u}_0,\overline{v}_0,\sigma_q,\sigma_u,{}&\sigma_v) = \\
&\frac{1}{(2 \pi)^{3/2}}
e^{-\frac{(\overline{q}-\overline{q}_0)^2 + (\overline{u}-\overline{u}_0)^2 + (\overline{v}-\overline{v}_0)^2}{2}},
\end{split}
\end{equation}
which is normalized, $\int_{-\infty}^\infty \int_{-\infty}^\infty \int_{-\infty}^\infty \overline{f}_C \, d\overline{q} \, d\overline{u} \, d\overline{v}=1$.
This distribution no longer explicitly depends on the $\sigma$'s and naturally handles
ellipsoidal distributions. The $\sigma$'s do, however, play a minor role: they limit
the possible values of $q_0$, $u_0$, and $v_0$ so it is important not to forget their
significance \citep{2012A&A...538A..65Q}.

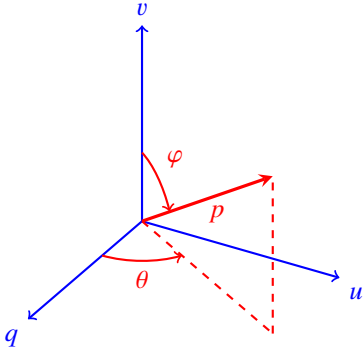
\begin{figure}[ht]
\tdplotsetmaincoords{60}{120}
\begin{tikzpicture}
	[scale=3,
		tdplot_main_coords,
		axis/.style={->,blue,thick},
		vector/.style={-stealth,red,very thick},
		vector guide/.style={dashed,red,thick},
		jason/.style={red,thin},
		angle/.style={red,thick}]

	\coordinate (O) at (0,0,0);
	
	\tdplotsetcoord{P}{1.4}{55}{60};
	\tdplotsetcoord{Q}{0.8}{55}{60};

	\draw[axis] (0,0,0) -- (1,0,0) node[anchor=north east]{$q$};
	\draw[axis] (0,0,0) -- (0,1,0) node[anchor=north west]{$u$};
	\draw[axis] (0,0,0) -- (0,0,1) node[anchor=south]{$v$};
	
	\draw[vector] (O) -- (P) ;
	\draw[jason] (O) -- (Q) node[anchor=north]{$p$};
	
	\draw[vector guide] (O) -- (Pxy);
	\draw[vector guide] (Pxy) -- (P);
	
	\tdplotdrawarc[angle,->]{(O)}{.35}{0}{60}{anchor=north}{$\theta$}

	\tdplotsetthetaplanecoords{55}
	
	\tdplotdrawarc[tdplot_rotated_coords,angle,->]{(O)}{.35}{0}{55}
          {anchor=south west}{$\varphi$}
\end{tikzpicture}
\caption{Illustration of the spherical coordinate system used in the paper.
The variable $\theta$ is used for the azimuthal angle with a range of $(-\pi,\pi]$,
and the variable $\varphi$ is measured from the positive $v$-axis with a
range of $[0,\pi]$.}
\label{coordsystem}
\end{figure}
We wish to now transform our equations to spherical coordinates as
the Poincar\'e sphere suggests. The transformation equations\footnote{In
this set of equations, the arctangent function, $\arctan(y, x)$, is the two-dimensional
version often
supported in computer languages. Its value is assumed to vary continuously
from $-\pi$ to $+\pi$
on the unit circle starting from the negative $x$-axis counterclockwise again to the negative $x$-axis.
Be extremely careful when implementing this function in computer code. Some
languages switch the arguments so that it is $\arctan(x, y)$.
Make sure your implementation gives valid angles for all quadrants
and axes,
especially if you use the single argument arctangent.} are
\begin{equation}
\begin{aligned}
p       ={}& \sqrt{q^2+u^2+v^2} \\
\theta  ={}& \arctan(u, q) \\
\varphi ={}& \arctan({\sqrt{q^2+u^2}}, v)
\end{aligned}
\end{equation}
which has inverse
\begin{equation}
\begin{aligned}
q ={}& p \sin(\varphi) \cos(\theta) \\
u ={}& p \sin(\varphi) \sin(\theta) \\
v ={}& p \cos(\varphi),
\end{aligned}
\end{equation}
where $\theta$ is the angle in the $q$-$u$ plane and $\varphi$ is
the angle from the positive $v$-axis (see Fig.~\ref{coordsystem}). The ``true values'',
$p_0$, $\theta_0$, and $\varphi_0$, are similarly
related to $q_0$, $u_0$, and $v_0$.

It is crucial not to confuse the degree of polarization, $p$,
with the degree of \textit{linear} polarization in the following. The degree of linear
polarization is a separate quantity
equal to $\sqrt{q^2+u^2}$ or $p \sin(\varphi)$ and,
although it is certainly important, is not discussed in this paper.

We now restrict ourselves to the case where all three
standard deviations are equal ($\Sigma_Q=\Sigma_U=\Sigma_V \equiv \Sigma$) so that $\sigma_q=\sigma_u=\sigma_v \equiv \sigma$. The distribution in spherical coordinates is then
\begin{equation}
\begin{split}
f'({}&p,\theta,\varphi|p_0,\theta_0,\varphi_0,\sigma) = \\
&\frac{p^2 \sin\varphi}{(2 \pi)^{3/2} \sigma^3}
e^{-\frac{p^2 + p_0^2 - 2 p p_0 \left( \sin\varphi \sin\varphi_0 \cos(\theta-\theta_0) + \cos\varphi \cos\varphi_0 \right)}{2 \sigma^2}}.
\end{split}
\label{fprime}
\end{equation}
The $p^2 \sin\varphi$ in the numerator of the factor out front is due to the Jacobian of the transformation, i.e., $dq \, du \, dv = p^2 \sin\varphi \, dp \, d\theta \, d\varphi$.\footnote{The function $f'$ is a scalar density and therefore gains a factor due to the
Jacobian under a coordinate transform.} It is rather
promising that the analytic form of Eq.~\ref{fprime} is
similar to and not much more complicated than the two-dimensional polar version.
Already one can see that the functional form of the
equation is not independent of the true value of the circular
polarization.

The barred version is
\begin{equation}
\begin{split}
\overline{f}'(\overline{p},{}&\theta,\varphi|\overline{p}_0,\theta_0,\varphi_0,\sigma) = \\
&\frac{\overline{p}^2 \sin\varphi}{(2 \pi)^{3/2}}
e^{-\frac{\overline{p}^2 + \overline{p}_0^2 - 2 \overline{p} \, \overline{p}_0 \left( \sin\varphi \sin\varphi_0 \cos(\theta-\theta_0) + \cos\varphi \cos\varphi_0 \right)}{2}},
\end{split}
\end{equation}
where $\overline{p} \equiv \sqrt{Q^2+U^2+V^2}/\Sigma = p / \sigma$ and $\overline{p}_0 \equiv \sqrt{Q_0^2+U_0^2+V_0^2}/\Sigma = p_0 / \sigma$.
This is still normalized, $\int_{0}^\pi \int_{-\pi}^\pi \int_0^{1/\sigma} \overline{f}' \, d\overline{p} \, d\theta \, d\varphi =1$.

\section{Classical estimators}
Now that $\overline{f}'(\overline{p},\theta,\varphi|\overline{p}_0,\theta_0,\varphi_0,\sigma)$
(the sampling distribution in signal-to-noise
variables) is at hand, two important classical estimators may be calculated.
These are the so-called ``Most Likely''
and ``Most Probable'' estimators. (Note that the
prime is not indicating a derivative in $\overline{f}'$ but serving
as a warning that we are using the $q$-$u$ plane angle, $\theta$, instead of
the sky angle for linear polarization.)

\subsection{The ``Most Likely'' solution}
The ``Most Likely'' (ML) solution is obtained by
maximizing $\overline{f}'$ with respect to $\overline{p}_0$, $\theta_0$, and $\varphi_0$.
Physically, this corresponds
to finding the value of $(\overline{p}_0, \theta_0, \varphi_0)$ 
which makes the measured value of $(\overline{p}, \theta, \varphi)$ the most
statistically likely, hence the name.
A general solution may be found by solving the system
$\frac{\partial \overline{f}'}{\partial \overline{p}_0}=0$,
$\frac{\partial \overline{f}'}{\partial \theta_0}=0$, and
$\frac{\partial \overline{f}'}{\partial \varphi_0}=0$ for
$\overline{p}_0$, $\theta_0$, and $\varphi_0$, which
are then treated as estimators. Technically,
a second derivative test must also be done to check that
the point is actually a maximum.
For functions of more than two variables, this test
requires checking that all the eigenvalues of the
function's Hessian matrix are positive. (All negative would
be a minima, while mixed values would be inconclusive.)
The equations involved in the test are very large
and cumbersome. From graphical investigation or intuition,
it is obvious, however, that the solution to be found is a maximum. 

From the azimuthal equation, $\frac{\partial \overline{f}'}{\partial \theta_0}=0$,
it is quickly deduced that $\theta_0=\theta$ everywhere except for values that
correspond to an origin (where $\overline{p}_0$ or $\overline{p}$ equals zero)
or a circular polarization axis (where $\varphi_0$ or $\varphi$ equals zero or $\pi$) of
the true and measured Poincar\'e spheres. Further tedious but straight-forward
algebra allows the full solution to be found, which is just 
\begin{equation}
\begin{aligned}
\overline{p}_{0,ML}={}& \overline{p} \\
\theta_{0,ML}={}& \theta\\
\varphi_{0,ML}={}& \varphi.
\end{aligned}
\label{MLsolution}
\end{equation}
This solution involves no ``bias corrections'' at all. What you
measure is your best estimate for the ``true'' polarization.

This solution is
consistent with the polar coordinate case (that is,
the two-dimensional case corresponding to $\varphi \rightarrow \pi/2$, $\varphi_0 \rightarrow \pi/2$, and
$\sigma_v \rightarrow 0$)
presented in \citet{2012A&A...538A..65Q}. It is only when the Rice distribution
(a one-dimensional distribution arising from
marginalization) is used that a non-trivial solution is found for
the ML method \citep{1985A&A...142..100S}. As marginalization intentionally removes
information present in the original distribution, the
logical basis for the use of non-trivial ML estimators
found from a marginalized distribution to perform
a ``bias correction'' is questionable. This is underscored
by the fact that the $p_0$, $\theta_0$, and $\varphi_0$
are not stochastic variables but input parameters in $\overline{f}'$ and
should not be treated as such as is done in the ML approach. The logically
correct way to ``invert'' the distribution is
to use Bayes' Theorem.

\subsection{The ``Most Probable'' solution}
The ``Most Probable'' (MP) solution is found by maximizing $\overline{f}'$ with respect
to $\overline{p}$, $\theta$, and $\varphi$.
This corresponds to finding
the $(\overline{p}_0,\theta_0,\varphi_0)$ point that
produces a distribution of measured points
with a maximum at the actual measured point $(\overline{p},\theta,\varphi)$.
A general solution may be found by solving the system
$\frac{\partial \overline{f}'}{\partial \overline{p}}=0$,
$\frac{\partial \overline{f}'}{\partial \theta}=0$, and
$\frac{\partial \overline{f}'}{\partial \varphi}=0$
for $\overline{p}_0$, $\theta_0$, and $\varphi_0$.
(As before, a second derivative test is technically
necessary to determine if the solution is actually
a maximum as opposed to a minimum or some point
of mixed inflection but this test is similarly impractical
to perform because the equations end up
being quite large. Graphically it can be shown
that the solutions are maximums.)

\begin{figure}
\includegraphics[width=3.5in]{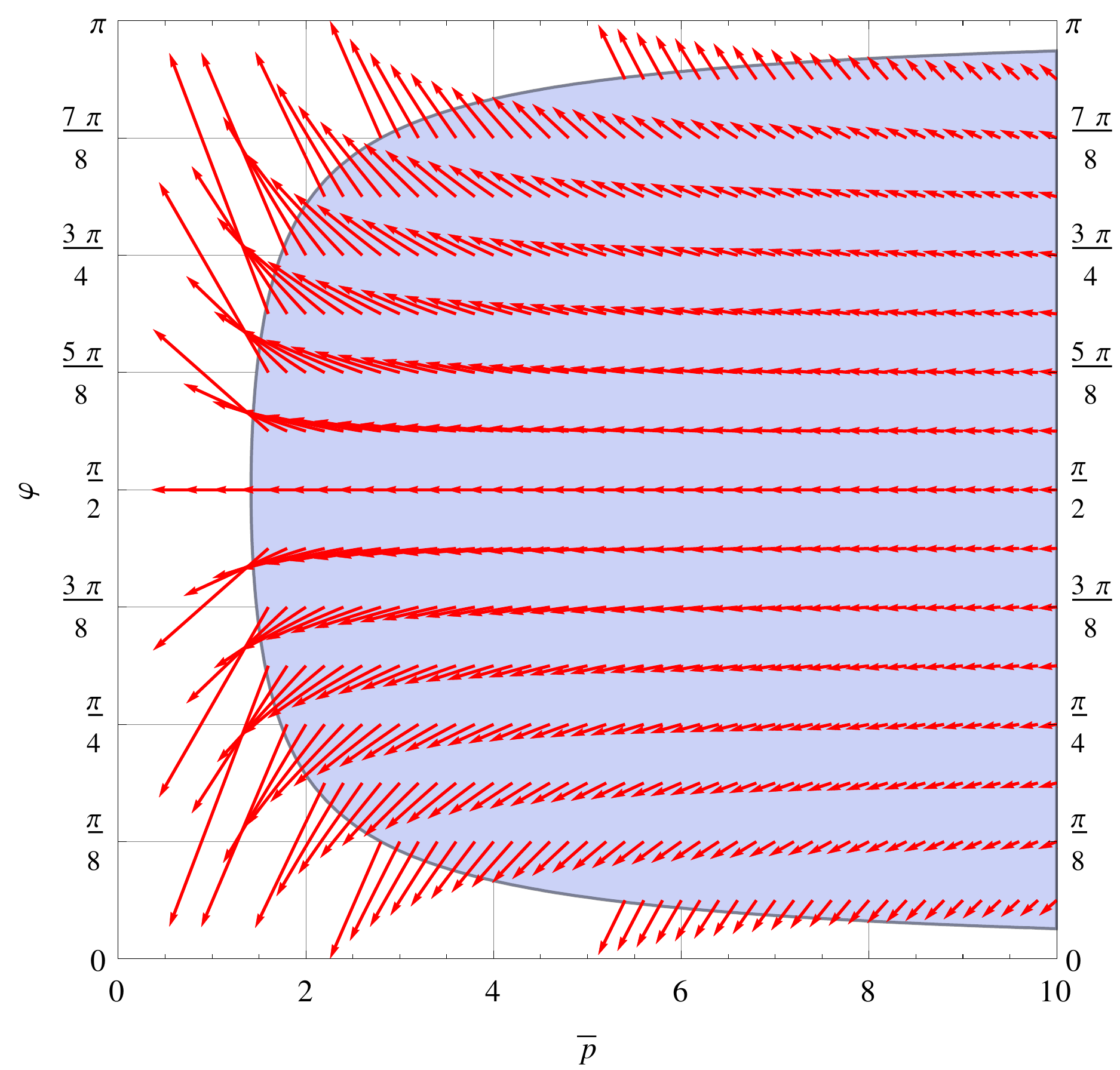}
\caption{Visualization of the $(\overline{p},\varphi)$
bias correction field of Eq.~\ref{MPsolution}. The tail of each arrow is located at a measured
$(\overline{p},\varphi)$ point and the head is attached to the
corresponding ``Most Probable'' estimate of $(\overline{p}_0,\varphi_0)$. This
correction should only be applied to $(\overline{p},\varphi)$ points lying
in the blue region given by Eq.~\ref{condition}. See the text for discussion
about $(\overline{p},\varphi)$ points outside the blue region.}
\label{biasfield}
\end{figure}

Again, one quickly finds $\theta_0=\theta$ from the azimuthal equation,
$\frac{\partial \overline{f}'}{\partial \theta}=0$. Using this
condition, the
$\varphi$ and $\overline{p}$ equations yield
$\overline{p}^2 = 2 + \overline{p} \, \overline{p}_0 \cos(\varphi - \varphi_0)$
and
$\overline{p} \, \overline{p}_0 \cos(\varphi_0) \sin(\varphi)^2 = \cos(\varphi) (1 + \overline{p} \, \overline{p}_0 \sin(\varphi) \sin(\varphi_0))$,
respectively. This two-equation system
is somewhat difficult to solve
and the solution is best accomplished with a computer algebra system.
The full analytic solution is
\begin{equation}
\begin{aligned}
\overline{p}_{0,MP} ={}& \frac{\sqrt{\cot(\varphi)^2+(\overline{p}^2-2)^2}}{\overline{p}}  \\
\theta_{0,MP} ={}& \theta \\
\varphi_{0,MP} ={}& \varphi - \arctan\left(\frac{\cot(\varphi)}{\overline{p}^2-2}\right),
\label{MPsolution}
\end{aligned}
\end{equation}
which is only valid for points
satisfying the condition
\begin{equation}
\overline{p} > \sqrt{1 + \csc(\varphi)^2},
\label{condition}
\end{equation}
where $\csc(x)$ is the cosecant function. Thus there are
two coupled ``bias corrections'' that must be made to the data:
one for $\overline{p}$ and one for $\varphi$. A graphical
representation of the preceding solution is given in Fig.~\ref{biasfield}.
The left-most extension of the region consisting of points
satisfying Eq.~\ref{condition} (shown in blue in the figure) approaches
the point $(\overline{p},\varphi)=(\sqrt{2},\pi/2)$. In the special
case of $\varphi=\pi/2$ and $\overline{p}>\sqrt{2}$, the correction
is just $\varphi_{0,MP}=\pi/2$ and $\overline{p}_{0,MP}=\overline{p}-2/\overline{p}$.
\begin{figure}
\includegraphics[width=3.5in]{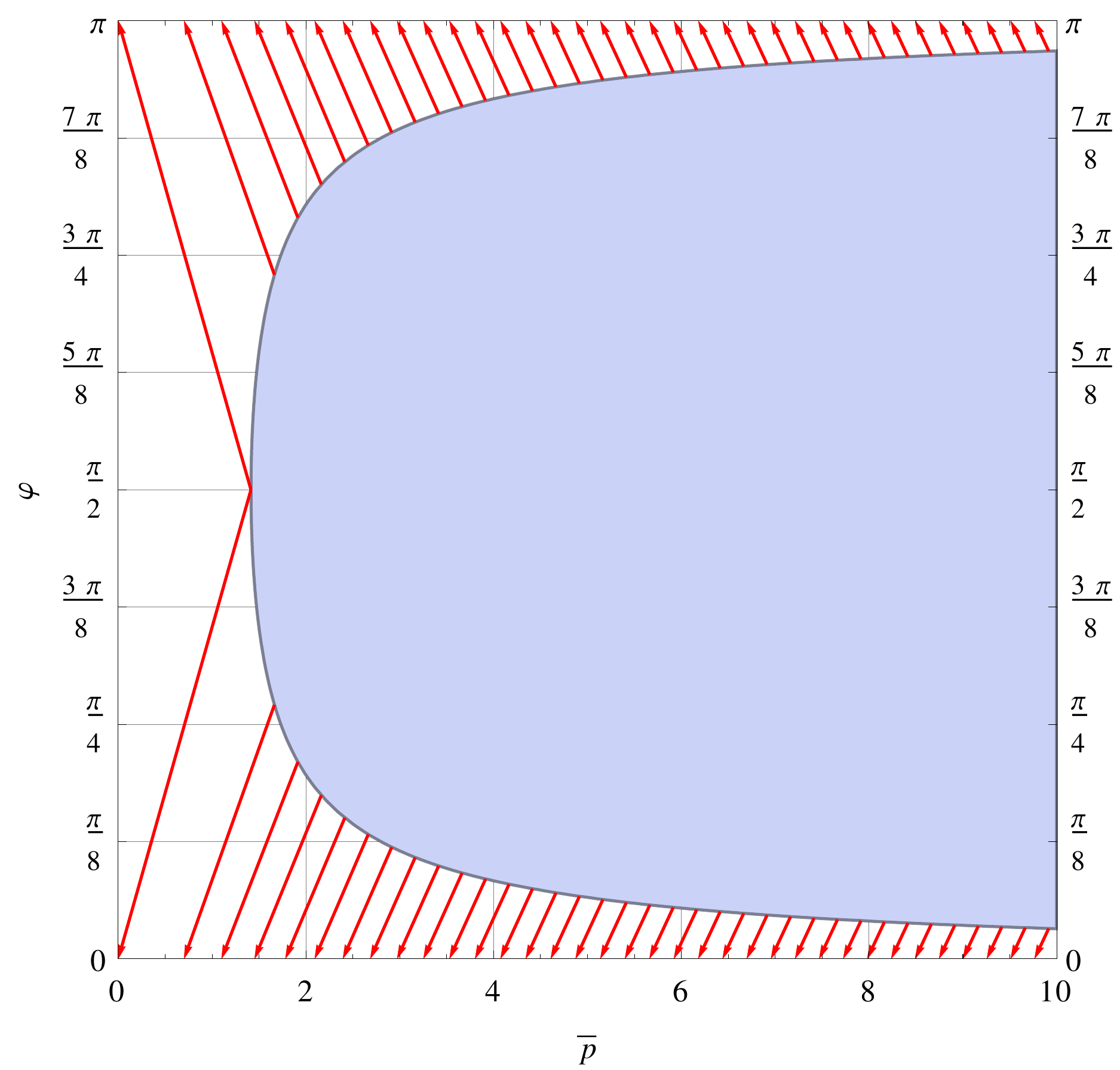}
\caption{Behavior of the bias correction field
(Eq.~\ref{MPsolution}) on the boundary of the blue valid region (Eq.~\ref{condition}).
The left-most point on the boundary of the blue region is at $(\sqrt2,\pi/2)$.}
\label{biasfield2}
\end{figure}

The ``bias correction field'' in Fig.~\ref{biasfield} shows
that the correction applied to $\overline{p}$
gets larger as $\overline{p}$ gets smaller (the actual
magnitude is $\varphi$-dependent).
This is consistent with the known behavior in the
two-dimensional case. Fig.~\ref{biasfield} also
shows that $\varphi$ itself must be corrected:
it should be made larger when $\varphi > \pi/2$
and smaller when $\varphi < \pi/2$. The
magnitude of this correction also tends to be
larger as $\overline{p}$ gets smaller. The
exact magnitude of the $\varphi$-correction at
a given $\overline{p}$ depends on $\varphi$ itself with the magnitude being
largest in the ``mid-latitudes'' of the Poincar\'e sphere
and zero on the equator. Generally, $(\overline{p},\varphi)$-points not on the equator are ``attracted'' towards the poles,
an effect most prominent at low signal-to-noise (between, say,
$\sqrt2 < \overline{p} \lesssim 5$). The
bias correction vector field diminishes as $\overline{p}$ gets larger
but for $\varphi$ values near the poles is seen to remain
significant even for $\overline{p}>10$, a regime typically
thought of as having signal-to-noise high enough not to have
to worry about such effects.

For points infinitesimally close to the boundary defined by
Eq.~\ref{condition}, the bias correction for $\varphi$ estimates
$\varphi_0=0$ or $\varphi_0=\pi$, whichever pole is nearest.
Fig.~\ref{biasfield2} illustrates this behavior. The special case of
$\varphi=\pi/2$ ``corrects'' to $\varphi_0=\pi/2$.
More specifically, points on the boundary with $\pi/2 < \varphi< \pi$ all
bias correct to $\varphi_0=\pi$ while those with $0 < \varphi< \pi/2$
all bias correct to $\varphi_0=0$. The $\overline{p}$ correction on the boundary acts
typically and corrects more and more towards the origin as $\overline{p}$ gets smaller.

From Figs.~\ref{biasfield} and \ref{biasfield2}, it can
be seen that every possible $(\overline{p}_0,\varphi_0)$
with $\overline{p}_0>0$ and $\sin \varphi_0 \ne 0$ is associated
with a $(\overline{p},\varphi)$ point within the blue region and vice
versa. Unfortunately, a closed form solution for the inverse of
Eq.~\ref{MPsolution} was not found and may not exist.

Intuitively,\footnote{This argument, while
``intuitive'' for understanding the bias correction
from a classical statistical perspective, can
easily be misapplied in a Bayesian scenario. Nevertheless,
when properly used, it is still an important tool to understand
Bayesian results. The key difference is that
in the classical approach, it applies to a single
``true'' point producing the measured points, whereas in the Bayesian approach
it applies individually to \textit{all} ``possibly
true'' points (weighted by their likelihood).} the $\overline{p}$ correction
may be understood by considering a $\overline{p}_0$ value
of zero. The measured value of $\overline{p}$ cannot be negative
and, due to scatter from measurement
error, you expect that $\overline{p}$ will almost always
be positive (that is, there is a negligibly small chance of zero)
even though $\overline{p}_0=0$. Thus, the measurement process ``biases'' 
$\overline{p}$ to larger values by some amount. Therefore, a correction
should be made that subtracts a bit from the measured
value to obtain a good estimate of $\overline{p}_0$.
Even for a non-zero $\overline{p}_0$, this biasing occurs although
with diminishing effect as $\overline{p}_0$ gets larger.
The same line of reasoning
can be used to gain an intuitive understanding of the
$\varphi$ bias correction. Consider a source
in a state of pure circular polarization. This corresponds
to a pole of the Poincar\'e sphere. Let's focus
on the purely right-circularly polarized case ($\varphi_0=0$).
Measurements of this state will be clustered around
the pole (including outside the Poincar\'e sphere)
but $\varphi$ is never negative and is almost
always positive.
Similarly for the left-circularly
polarized case ($\varphi_0=\pi$), measurements of $\varphi$ will be
clustered near that pole but cannot have $\varphi$ greater 
than $\pi$ and almost always will have $\varphi$ less than $\pi$. Just as with $\overline{p}$, a
correction is needed to remove this effect from $\varphi$. It is important
to remark that both the $\overline{p}$ and $\varphi$ bias corrections
are purely coordinate effects. It may seem at first as if
the $\varphi$-correction is inconsistent with the symmetry of the
system (from $\Sigma_Q=\Sigma_U=\Sigma_V$) but it is not. While conventionally the $v$-axis
is always chosen to isolate circular polarization, the effect would arise
regardless of the orientation of the $v$-axis within
the Poincar\'e sphere, which preserves the overall symmetry (i.e., the
symmetry is spontaneously broken by introduction of the coordinate
system).

It must be stressed that the solution of Eq.~\ref{MPsolution} only applies
to measured points satisfying Eq.~\ref{condition} (illustrated
by the blue region in the figures). Presumably, the area outside
the blue region divides into three other regions:
one characterized by $\overline{p}_{0,MP}=0$, another by
$\overline{p}_{0,MP} \ne 0$ and $\varphi_{0,MP}=0$,
and still another by $\overline{p}_{0,MP} \ne 0$ and $\varphi_{0,MP}=\pi$. In the latter two cases, there
could still be a $\overline{p}$ bias correction at work.
It is unclear how to determine mathematically a ``MP solution'' in the non-blue region
of Fig.~\ref{biasfield}. A formal
solution may simply not exist there.
The authors' best guess as to the best
practice is that measurements
within the ``triangle'' defined by points
$(0,0)$, $(0,\pi)$, and $(\sqrt{2},\pi/2)$
correct to $(0,\pi)$ if $\varphi>\pi/2$, $(0,0)$ if $\varphi<\pi/2$, and
$(0,\pi/2)$ if $\varphi=\pi/2$.\footnote{It may seem strange
to worry about the $\varphi$ angle's best estimate if $\overline{p}_{0,MP}=0$
but using a ``proper'' solution may have nicer continuity properties and
be better for computer programs than, for instance, simply assuming or defining $\varphi_{0,MP}=\pi/2$.} The remaining points could 
correct to the same value as the blue region boundary point whose
``correction arrow'' passes through the point.
Another possibility for these later points is
to translate the correction arrow
from the boundary of the blue region with the same $\varphi$
to the point and use the intersection of that arrow
with the $\overline{p}$-axis as the corrected value.
Formal solutions in these regions where not, however, found.

Eq.~\ref{MPsolution} does indeed provide an
estimate of the $(\overline{p}_0,\varphi_0)$ point that
would produce a distribution of measured points
with maximum at $(\overline{p},\varphi)$. It is however
just a point estimate. The actual $(\overline{p},\varphi)$
distribution for a small values of $\overline{p}_0$ tends to be broad, shallow,
and nearly flat, especially in the $\varphi$ dimension. In other words,
the errors bars associated with each estimate will be large
so it is cautioned that the point
estimate should not be taken too literally.
Using the $\overline{f}'$ distribution, one could 
construct confidence regions. 
These regions are however not uniquely determined, even
in the one-dimensional case \citep{1985A&A...142..100S}. Construction of confidence
regions is reserved for future work.

\section{Marginalized distributions}
Experimenters often work with only a partial
description of a physical system. This
may be necessary when independent variables of a distribution
of interest related to the system are not measurable,
or desired when there are so-called nuisance variables
of little concern. In a polarization study, for instance, it may be that only the degree
of linear polarization is the focus and not the angle of linear
polarization. In such situations, one
can marginalize over ``unwanted'' independent variables to find a new
distribution which no longer depends on them
but at the cost of losing some information present in the
original distribution.

The marginalization possibilities for $f'$ are to marginalize over
\begin{inparaenum}[(I)]
  \item $p$ only;\label{p}
  \item $\theta$ only;\label{t}
  \item $\varphi$ only;\label{f}
  \item $p$ and $\theta$;\label{pt}
  \item $p$ and $\varphi$; and\label{pf}
  \item $\theta$ and $\varphi$.\label{tf}
\end{inparaenum} The first and last cases are perhaps the most interesting
and shall now be investigated.

\subsection{The $p$-marginalized angular distribution}
The $p$-marginalized angular distribution is
\begin{equation}
M(\theta, \varphi|p_0,\theta_0,\varphi_0,\sigma) = \int_{0}^{\infty} f'(p,\theta,\varphi|p_0,\theta_0,\varphi_0,\sigma) \, dp
\end{equation}
or, in signal-to-noise variables,
\begin{equation}
\overline{M}(\theta, \varphi|\overline{p}_0,\theta_0,\varphi_0,\sigma) = \int_{0}^{\infty} \overline{f}'(\overline{p},\theta,\varphi|\overline{p}_0,\theta_0,\varphi_0,\sigma) \, d\overline{p}.
\end{equation}
This equation can be simplified. Start with
\begin{equation*}
\begin{aligned}
\overline{M}&(\theta, \varphi|\overline{p}_0,\theta_0,\varphi_0,\sigma) \\
=& \int_{0}^{\infty}
\frac{\overline{p}^2 \sin\varphi}{(2 \pi)^{3/2}}
e^{-\frac{\overline{p}^2 + \overline{p}_0^2 - 2 \overline{p} \, \overline{p}_0 \left( \sin\varphi \sin\varphi_0 \cos(\theta-\theta_0) + \cos\varphi \cos\varphi_0 \right)}{2}}
d\overline{p} \\
=&
\frac{\sin\varphi}{(2 \pi)^{3/2}}
e^{-\frac{\overline{p}_0^2}{2}}
\int_{0}^{\infty}
\overline{p}^2 
e^{-\frac{\overline{p}^2 - 2 \overline{p} \, \overline{p}_0 \left( \sin\varphi \sin\varphi_0 \cos(\theta-\theta_0) + \cos\varphi \cos\varphi_0 \right)}{2}}
d\overline{p}
\end{aligned}
\end{equation*}
and use the definite integral
\begin{equation}
\int_0^\infty \! x^2 e^{-\frac{x^2+a x}{2}} dx = \sqrt{\frac{\pi}{32}} (a^2+4) e^{a^2/8} \operatorname{erfc}\left( \frac{a}{2\sqrt{2}} \right) -\frac{a}{2},
\end{equation}
where $\operatorname{erfc}(x)$ is the complementary error function defined as\footnote{Some authors define $\operatorname{erfc}(x)$ without the $\frac{2}{\sqrt{\pi}}$ factor.}
\begin{equation}
\operatorname{erfc}(x) \equiv 1 - \operatorname{erf}(x) = \frac{2}{\sqrt{\pi}} \int_x^\infty e^{-t^2} dt
\end{equation}
and $\operatorname{erf}(x)$ is the error function to reduce this integral.
\begin{equation}
\begin{aligned}
\overline{M}(\theta, \varphi|\overline{p}_0,\theta_0, \varphi_0,{}& \sigma) =  \\
\frac{\sin\varphi}{(2 \pi)^{3/2}} 
e^{-\frac{\overline{p}_0^2}{2}} &\left( \sqrt{\frac{\pi}{32}} (A^2+4) e^{A^2/8} \operatorname{erfc} \left( \frac{A}{2\sqrt{2}} \right) -\frac{A}{2} \right)
\label{pmarginalized}
\end{aligned}
\end{equation}
is obtained where 
\begin{equation}
A \equiv -2 \, \overline{p}_0 \left( \sin\varphi \sin\varphi_0 \cos(\theta-\theta_0) + \cos\varphi \cos\varphi_0 \right)
\end{equation}
which may be viewed as a higher-dimensional analog of the angular
distribution presented in \citet{1965AnAp...28..412V} (see also \citet{1986VA.....29...27C,1993A&A...274..968N,2012A&A...538A..65Q}).

Examining a few density plots of Eq.~\ref{pmarginalized} for various values of $\overline{p}_0$, $\theta_0$, and $\varphi_0$ reveals that the probability distribution is attenuated near the poles of the coordinate system, an effect that becomes broader for small values of $\overline{p}_0$. In particular, one should notice that even for $\overline{p}_0=0$, the probability density still varies with $\varphi$. This is due to the pole bias effect discussed earlier.

One would like to go further and find the two one-dimensional marginalized angular distributions (cases
IV and V) but the integrations seem unable to be completely performed using elementary or even common special functions.

\subsection{The angular-marginalized $p$-distribution}
One is often primarily concerned with the degree
of polarization, $p$, and not the value of the
angular variables. Marginalization over
the angular variables will therefore produce
a new distribution, to be called
$\myletter$ ($\overline{\myletter}$ in signal-to-noise variables)
of high practical importance. In the polar coordinate case, marginalization
over the (lone) angular variable, $\theta$, produces
the Rice distribution. In the spherical
coordinate case, the marginalization must occur
over both $\theta$ and $\varphi$.
Thus, the function $\overline{\myletter}$ may be considered
a higher-dimensional analog of the Rice distribution.
Let
\begin{equation}
\myletter(p|p_0,\theta_0,\varphi_0,\sigma) = \int_{0}^{\pi} \! \int_{-\pi}^{\pi} f'(p,\theta,\varphi|p_0,\theta_0,\varphi_0,\sigma) \, d\theta \, d\varphi
\end{equation}
or similarly
\begin{equation}
\overline{\myletter}(\overline{p}|\overline{p}_0,\theta_0,\varphi_0,\sigma) = \int_{0}^{\pi} \! \int_{-\pi}^{\pi} \overline{f}'(\overline{p},\theta,\varphi|\overline{p}_0,\theta_0,\varphi_0,\sigma) \, d\theta \, d\varphi.
\end{equation}

Let's calculate.
\begin{align}
\begin{aligned}
\overline{\myletter}{}&(\overline{p}|\overline{p}_0, \theta_0,\varphi_0,\sigma) = \frac{\overline{p}^2}{(2 \pi)^{3/2}}
e^{-\frac{\overline{p}^2+\overline{p}_0^2}{2}} \times \\
 &\int_{0}^{\pi} \! \int_{-\pi}^{\pi} \!
\sin\varphi \, e^{\overline{p} \, \overline{p}_0 ( \sin\varphi \sin\varphi_0 \cos(\theta-\theta_0) + \cos\varphi \cos\varphi_0)}
\, d\theta \, d\varphi \\
={} & \frac{\overline{p}^2}{(2 \pi)^{3/2}}
e^{-\frac{\overline{p}^2+\overline{p}_0^2}{2}} \times \\
 & \int_{0}^{\pi} \!
\sin\varphi \, e^{\overline{p} \, \overline{p}_0 \cos\varphi \cos\varphi_0 }
\int_{-\pi}^{\pi} \!
e^{\overline{p} \, \overline{p}_0 \sin\varphi \sin\varphi_0 \cos(\theta-\theta_0) }
\, d\theta \, d\varphi 
\end{aligned}
\end{align}
The $\theta$-integral can be solved using 
$\int_{-\pi}^{\pi} e^{a \cos(t)} dt = 2\pi \mathcal{I}_0(a)$,
where $\mathcal{I}_0(x)$ is the modified Bessel function of the first kind.
$\mathcal{I}_0$ is an even, real function when its argument is real.
(Because the range of integration is over a full period of $\theta$, the value of
$\theta_0$ is immaterial because it appears as
a phase shift and does not affect the integral.) Once
the $\theta$-integral is performed,
the $\overline{\myletter}$ distribution no longer explicitly
depends on $\theta_0$.
We shall therefore drop $\theta_0$ from the conditions.
Continuing with the calculation,
\begin{align}
\begin{aligned}
\overline{\myletter}(\overline{p}|&\overline{p}_0,\varphi_0,\sigma) = \frac{\overline{p}^2}{(2 \pi)^{3/2}}
e^{-\frac{\overline{p}^2+\overline{p}_0^2}{2}} \times \\
& \int_{0}^{\pi} \!
\sin\varphi \, e^{\overline{p} \, \overline{p}_0 \cos\varphi \cos\varphi_0 }
(2 \pi) \mathcal{I}_0(\overline{p} \, \overline{p}_0 \sin\varphi \sin\varphi_0)
d\varphi \\
={}& \frac{\overline{p}^2}{\sqrt{2 \pi}}
e^{-\frac{\overline{p}^2+\overline{p}_0^2}{2}} \times \\
&\int_{0}^{\pi} \!
\sin\varphi \, e^{\overline{p} \, \overline{p}_0 \cos\varphi \cos\varphi_0 }
\mathcal{I}_0(\overline{p} \, \overline{p}_0 \sin\varphi \sin\varphi_0)
d\varphi.
\label{bigbad}
\end{aligned}
\end{align}

The $\varphi$-integral in the last expression is non-trivial
but it appears it can be solved. Let 
\begin{equation}
\omega(a,\varphi,\varphi_0) \equiv \sin\varphi \, e^{a \cos\varphi \cos\varphi_0} \mathcal{I}_0(a \sin\varphi \sin\varphi_0)
\label{hardfunc}
\end{equation}
and
\begin{equation}
\Omega(a,\varphi_0) \equiv \int_{0}^{\pi} \! \omega(a,\varphi,\varphi_0) d\varphi,
\label{hardint}
\end{equation}
where $a \equiv \overline{p} \, \overline{p}_0 \ge 0$.
No computer algebra system tested could solve the integral in
Eq.~\ref{hardint} and no applicable integrals
were found in tables of integrals. Manual attempts to solve it also failed.
From graphs of $\Omega$ versus $\varphi_0$ created numerically\footnote{This is a sensitive result and may require use of a program
or library that allows the user to specify a higher level of precision than the native floating point size.} for several values of $a$, it is noticed that the function may actually be constant in $\varphi_0$ for a given $a$. 
 Unfortunately,
it is not a simple matter to show $\frac{\partial\Omega}{\partial\varphi_0} = 0$ 
but if $\frac{\partial\Omega}{\partial\varphi_0}$ is graphed for $0 \le \varphi_0 \le \pi$ at several sampled values of $a$ it seems to always show residuals about zero with scatter consistent with numerical
noise. The previous numerical observations lead one to suspect very strongly that $\Omega(a,\varphi_0)$ is independent of the value of $\varphi_0$. 
This may be surprising in light of the bias corrections previously discussed;
however, the same marginalized formula seems like it should be found regardless of the rotational orientation
of the $(\overline{q},\overline{u},\overline{v})$ coordinate system within the
Poincar\'e sphere. 
If it is true that $\Omega(a,\varphi_0)$ is constant in $\varphi_0$, any value of $\varphi_0$ may be
chosen and it does not alter $\Omega(a,\varphi_0)$'s value at a given $a$. In that case,
a value of $\varphi_0$ may be judiciously chosen that simplifies the
integral. Good choices are $\varphi_0=0$ or $\pi$ for which
\begin{equation}
\Omega(a,\varphi_0) = \frac{2 \sinh(a)}{a}
\label{unprovedresult}
\end{equation}
is found using the common integral $\int_0^\pi \! e^{x \cos{\varphi}} \sin{\varphi} \, d\varphi = 2\sinh(x)/x$, where $\sinh(x)$ is the hyperbolic sine of $x$. Indeed, simple
arguments modifying the factors in
the integrand of Eq.~\ref{hardint} can prove this to be a lower limit
for $\Omega(a,\varphi_0)$ at a given $a$ (see Appendix~\ref{boundappendix}). Proof that it is also the upper limit was elusive (except at $\varphi_0=0$, $\pi/2$, and $\pi$).
A Taylor series expansion of $\Omega(a,\varphi_0)$ in $\varphi_0$ at $\varphi_0=0$ was found to be consistent with Eq.~\ref{unprovedresult} out to 26-th order using computer algebra systems.\footnote{
The zeroth-order term of the expansion is $2 \sinh(a)/a$. If Eq.~\ref{unprovedresult}
is true, then it is the only non-zero term. It was proven that
all the odd-order coefficients in the expansion are identically zero.
It appears that the positive even-order expansion coefficients will also always be
identically zero but this remains unproven.}
Based on physical suspicion corroborated by numerical results,
Eq.~\ref{unprovedresult} shall be assumed to be true henceforth.

Returning to the main distribution and using Eq.~\ref{unprovedresult}, $\overline{\myletter}$, it is proposed that
\begin{equation}
\overline{\myletter}(\overline{p}|\overline{p}_0,\sigma) = \sqrt{\frac{2}{\pi}} \frac{\overline{p}}{\overline{p}_0} e^{-\frac{\overline{p}^2+\overline{p}_0^2}{2}} \sinh(\overline{p} \, \overline{p}_0).
\label{mydistribution}
\end{equation}
Since the value of $\varphi_0$ is irrelevant,
$\overline{\myletter}$'s dependence on $\varphi_0$ has been dropped from the notation.
This is the higher-dimensional analog of the Rice distribution.
Surprisingly, it is a simpler function than the Rice distribution
in the sense that it depends on the hyperbolic sine function rather
than a modified Bessel function.
The $\overline{\myletter}$ distribution is
normalized, $\int_0^\infty \! \overline{\myletter} \, d\overline{p} =1$.
\begin{figure}
\includegraphics[width=3.5in]{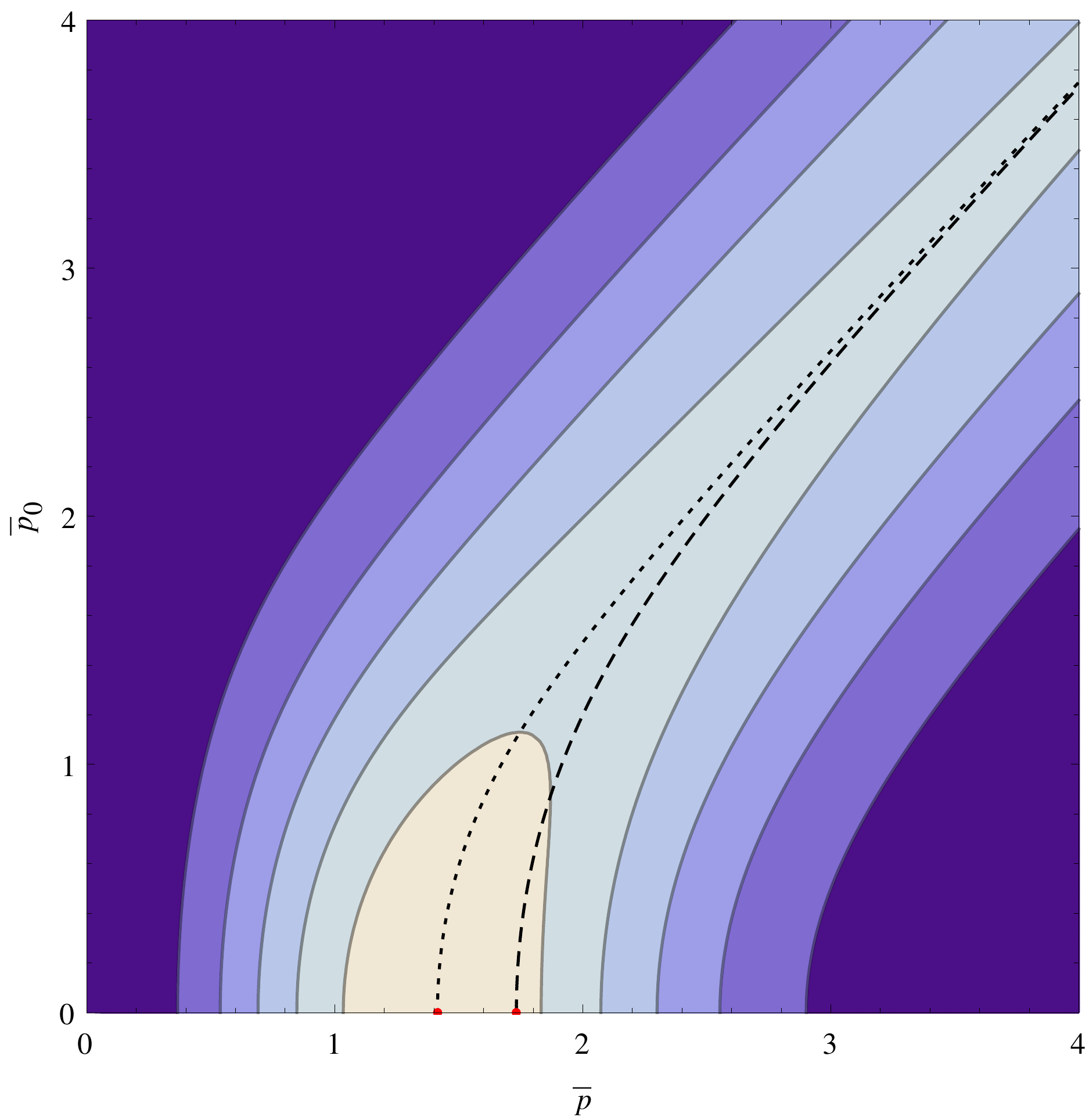}
\caption{Contour plot of the angular-marginalized $\overline{p}$-distribution
given by Eq.~\ref{mydistribution}.
The contours are at $0.1$ intervals with the lighter shades
corresponding to larger values. The dotted and dashed curves (Eqs.~\ref{hslices} and \ref{vslices})
trace maximums along horizontal and vertical slices and end at
the red points at $(\sqrt{2},0)$ and $(\sqrt{3},0)$, respectively.
Although the ``shape'' of the distribution is not affected by the
value of $\sigma$, the plot assumes that $\sigma \le 1/4$
so that the plotted values for $\overline{p}_0$ up to $4$ ($=1/\sigma_{max}$) exist.}
\label{quinnplot}
\end{figure}

Some general properties of $\overline{\myletter}$ can now be investigated.
A contour plot of the function is shown in Fig.~\ref{quinnplot} with
contours at $0.1$ intervals. The function
achieves a global maximum 
as $(\overline{p},\overline{p}_0)$ approaches $(\sqrt{2},0)$ of $\frac{2}{e} \sqrt{\frac{2}{\pi}} \approx 0.587051$.
Tracing the maximums of horizontal slices (the dotted line in the figure) yields the equation
$(\overline{p}^2-1) \sinh(\overline{p} \, \overline{p}_0) - \overline{p} \, \overline{p}_0 \cosh(\overline{p} \, \overline{p}_0)=0$
or
\begin{equation}
(\overline{p}_0^2-1) \tanh(\overline{p} \, \overline{p}_0) = \overline{p} \, \overline{p}_0
\label{hslices}
\end{equation}
This implicit curve intersects the $\overline{p}$-axis at the global maximum at $(\sqrt{2},0)$ and
is the ``Most Probable''
estimator curve for $\overline{\myletter}$.
Tracing the maximums of vertical slices (dashed line) yields
$\overline{p} \, \overline{p}_0 \cosh(\overline{p} \, \overline{p}_0) - (\overline{p}_0^2+1) \sinh(\overline{p} \, \overline{p}_0)=0$
or
\begin{equation}
(\overline{p}_0^2+1) \tanh(\overline{p} \, \overline{p}_0) = \overline{p} \, \overline{p}_0
\label{vslices}
\end{equation}
This curve intersects at the $\overline{p}$-axis at $(\sqrt{3},0)$ and
is the ``Most Likely'' estimator curve for $\overline{\myletter}$.

The MP estimator curve of Eq.~\ref{hslices} (dotted
curve in Fig.~\ref{quinnplot}) can be used for all
$\overline{p}>\sqrt2$. Yet it was stressed
that the full MP
solution given Eqs.~\ref{MPsolution} and \ref{condition}
is only valid for some values with $\overline{p}>\sqrt2$.
Marginalization has thus hidden some of the complexity of the problem.
The ML estimator given by Eq.~\ref{vslices}
(dashed curve) now has non-trivial behavior unlike the full
solution in Eq.~\ref{MLsolution}. It is not easy to anticipate
the effect that marginalization will have
on various estimators.

It is interesting to compare the long form of these estimator curves and the
corresponding curves based on the Rice
distribution \citep{1974ApJ...194..249W,1985A&A...142..100S,2012A&A...538A..65Q}.
These curves are based on the hyperbolic trigonometric
functions whereas the curves in the two-dimensional case
are based on Bessel functions.
Their points of intersection
with the $\overline{p}$-axis have also been shifted by $1$. The differences
arise because the Jacobian factor is proportional to $p^2$ in
spherical coordinates but only $p$ for polar coordinates.

\section{Discussion}
How do these new results
relate to the old results derived using polar
coordinates for the ``Poincar\'e disk''? First, the old results are \textit{not}
special cases of the new results when
$\varphi \rightarrow \pi/2$ and $\varphi_0 \rightarrow \pi/2$;
for instance, in the special case of $\varphi=\pi/2$, the
$\overline{p}$ equation of Eq.~\ref{MPsolution}
reduces to $\overline{p}-2/\overline{p}$ rather than the
Wang estimator $\overline{p}-1/\overline{p}$ \citep{1997ApJ...476L..27W,2012A&A...538A..65Q}
and $\overline{M}(\theta,\varphi)$ does not reduce to angular
distribution given by \citet{1965AnAp...28..412V}.
Similarly, $\overline{\myletter}$ is altogether distinct from
the Rice distribution. The new results
are therefore not direct extensions of the old functions
but they do have somewhat similar forms.
The three versus two dimensional nature of the parameter
space is responsible for this.

Particularly interesting is a comparison of Fig.~\ref{quinnplot}
with fig.~{2a} of \citet{2012A&A...538A..65Q}. The
intercepts of all the estimator curves have undergone shifts
of the form $\sqrt{n} \rightarrow \sqrt{n+1}$. This again
is due to the increased dimensionality of the problem.
Thus, when circular polarization has equal measurement error
to the linear Stokes axes' measurement error,
it takes a bigger measurement of
the magnitude of the polarization vector to be significant
than the two-dimensional case suggests.
The two-dimensional
results can perhaps be recovered by some limiting procedure
where $\sigma_v \rightarrow 0$
while $\sigma$ ($=\sigma_q=\sigma_u$) is held constant.
This suggests that decreasing the measurement error on
one of $\sigma_q$, $\sigma_u$, or $\sigma_v$ should
decrease the $\overline{p}$-threshold for a significant measurement.
On the other hand, often an instrument measures only linear or only circular
polarization. This may correspond to a scenario where
the distribution of Stokes parameters is best modeled
by assuming a flat (instead of gaussian) distribution
for the variables of the undetected type. This would
violate our starting assumptions in Eq.~\ref{starteq}
and would produce rather different results. As
a flat distribution is in an informal sense a ``wide''
distribution, it is probable that this would raise the detection
threshold higher.
It should be clear that interpreting polarization
data at low signal-to-noise in ``human digestible'' quantities
like degrees of polarization or angle of polarization
is difficult.

\section{Conclusion}
The three-dimensional Stokes sampling distribution in spherical
coordinates is given. From it the ``Most Likely''
and ``Most Probable'' classical (that is, non-Bayesian) estimators 
for the ``true'' Stokes parameters given
measured values have been derived. Additionally, a two useful marginalizations
of the sampling distribution have been calculated including a higher
dimensional analog of the Rice distribution.
These results are necessary stepping stones towards
a full and proper treatment of polarization measurements.

It is cautioned that a deep understanding of the measurement
of Stokes parameters in spherical coordinates like
here or the polar coordinates usually used for
linear polarization requires a Bayesian approach. The presented results may
nonetheless have tolerable accuracy for many experiments.

While the setup of the problem and the
presentation of the results has been
framed in terms of the study of polarization, the
solutions are in fact far more general. They
are applicable to the measurement of any ``source point''
contained in a three-dimensional ball
measured with Gaussian error in each Cartesian direction and
presented in spherical coordinates. The results could be used to study the location
of neutrino production within the Sun or
the determination of the epicenter
of an earthquake. The results could also influence
the interpretation of cosmic microwave background
polarization data.
Extensions of the work removing the constraint
that the error be equal in each direction
or that involve correlated variables
would be highly desirable.

\begin{appendix}

\section{Bounds on $\Omega(a,\varphi_0)$}
\label{boundappendix}
It helps to have functional bounds on
\begin{equation}
\Omega(a,\varphi_0) \equiv \int_0^\pi \! \sin(\varphi) e^{a \cos(\varphi) \cos(\varphi_0)} I_0(a \sin(\varphi) \sin(\varphi_0)) \, d\varphi
\label{origeq}
\end{equation}
to show, for instance, that it is at least non-infinite over the
$\varphi_0$-domain of interest.
There are three factors ($\sin(\varphi)$, $e^{a \cos(\varphi) \cos(\varphi_0)}$, and $I_0(a \sin(\varphi) \sin(\varphi_0))$) in the integrand,
each of which is simple enough that they allow
various bounds to be obtained for the total expression.

\subsection{Lower bound}

Under variation of $\varphi_0$, the $e^{a \cos(\varphi) \cos(\varphi_0)}$ factor is minimized when $\cos(\varphi_0)$ is the smallest, which occurs for $\varphi_0=\pi$, so that the factor
becomes $e^{-a \cos(\varphi)}$. Furthermore, the $I_0(a \sin(\varphi) \sin(\varphi_0))$ factor is the smallest
when its argument is zero, which also occurs at $\varphi_0=\pi$ (also
$\varphi_0=0$) so that the factor become $\mathcal{I}_0(0)=1$. After these
substitutions, the integral for $\Omega$ can be performed explicitly:
\begin{equation}
\int_0^\pi \! \sin(\varphi) \, e^{-a \cos(\varphi)} d\varphi = \frac{2 \sinh(a)}{a}.
\end{equation}
For any given value of $a$, this sets a (constant) global lower bound on the integral: the value of integral
must be greater than or equal to $\frac{2 \sinh(a)}{a}$ for all values of $\varphi_0$.
It is probably also the greatest lower bound for each $\varphi_0$.

\subsection{Upper bounds}
The $e^{a \cos(\varphi) \cos(\varphi_0)}$ factor is maximized when its exponent is the largest while the $I_0(a \sin(\varphi) \sin(\varphi_0))$ factor is the largest
when the magnitude of its argument is as big as possible.

A constant global upper bound is found by modifying the factors in the
obvious way:
\begin{equation}
\int_0^\pi \! \sin(\varphi) \, e^{a \cos(\varphi)} \mathcal{I}_0(\pm a) \, d\varphi = \frac{2 \sinh (a) }{a} \mathcal{I}_0(a).
\end{equation}
A stricter (but non-constant) global upper bound can
be found merely by allowing $\sin(\varphi_0) \rightarrow 1$ in
the Bessel function factor,
\begin{equation}
\begin{aligned}
\int_0^\pi \! \sin(\varphi)& \, e^{a \cos(\varphi) \cos(\varphi_0)} \mathcal{I}_0(a \sin(\varphi)) \, d\varphi \\
&= \frac{2 \sec(\varphi_0) \sinh (a \cos(\varphi_0))}{a} \mathcal{I}_0(a)   \quad \quad \text{for } \varphi_0 \ne \pi/2.
\end{aligned}
\label{bestbound}
\end{equation}
This last solution has functional dependence on $\varphi_0$ and is not valid
for $\varphi_0=\pi/2$ because of the $\sec(\varphi_0)$ component. For $\varphi_0=0$
this bound is $\frac{2 \sinh (a) }{a} \mathcal{I}_0(a)$ and it decreases smoothly and approaches
$\frac{2 \sinh(a)}{a}$ as $\varphi_0 \rightarrow \pi/2^+$. Similarly, for $\varphi_0=\pi$ it
is $\frac{2 \sinh (a) }{a} \mathcal{I}_0(a)$ and it decreases smoothly and approaches
$\frac{2 \sinh(a)}{a}$ as $\varphi_0 \rightarrow \pi/2^-$. Of course,
if $\varphi_0=\pi/2$, Eq.~\ref{origeq} can be accomplished directly:
\begin{equation}
\int_0^\pi \sin(\varphi) \, e^{0} \mathcal{I}_0(a \sin(\varphi)) \, d\varphi = \frac{2 \sinh(a)}{a} \quad \text{for } \varphi_0=\pi/2,
\end{equation}
so the discontinuity in Eq.~\ref{bestbound} at $\varphi_0=\pi/2$ can be ``removed''. Eq.~\ref{origeq} can also be performed for $\varphi_0=0$ and $\varphi_0=\pi$ and in both
cases $\frac{2 \sinh(a)}{a}$ results, 
so we have explicit upper bounds at those values too.

\subsection{Best bounds}
Combining the previous information, the best bounds are
\begin{equation}
\frac{2 \sinh(a)}{a} \le \Omega(a,\varphi_0) \le \begin{cases} \frac{2 \sinh(a)}{a} &\mbox{if } \varphi_0=0,\frac{\pi}{2},\pi  \\   \frac{2 \sec(\varphi_0) \sinh (a \cos(\varphi_0))}{a} \mathcal{I}_0(a) &\mbox{otherwise,} \end{cases}
\end{equation}
which are finite for all $\varphi_0$ at a given value of $a$.

\end{appendix}

\bibliographystyle{aa}
\bibliography{stokes3d}

\end{document}